\documentclass{article}
\usepackage[T1]{fontenc} 
\usepackage[utf8]{inputenc} 
\usepackage{ismir,amsmath,cite,url, amssymb}
\RequirePackage[linesnumbered,ruled]{algorithm2e}
\usepackage{color}
\usepackage{multirow}
\usepackage{booktabs}
\usepackage{pgffor}
\usepackage{amsfonts}
\usepackage{subcaption}
\usepackage{amssymb}
\usepackage[pdftex]{graphicx}

\usepackage{lineno}

\title{A Semi-Supervised Deep Learning Approach to Dataset Collection for Query-by-Humming Task}
\multauthor
{Amantur Amatov$^1$ \hspace{1cm} Dmitry Lamanov$^2$ \hspace{1cm} Maksim Titov$^2$} { \bfseries{Ivan Vovk$^2$ Ilya Makarov$^3$ Mikhail Kudinov$^2$}\\
 $^1$ Higher School of Economics, Moscow, Russia\\
$^2$ Huawei Noah's Ark Lab, Moscow, Russia\\
$^3$  AI Center, NUST MISiS, Moscow, Russia
}
\def\authorname{A. Amatov, D. Lamanov, M. Titov, I. Vovk, I. Makarov, M. Kudinov}
\usepackage[bookmarks=false,pdfauthor={\authorname},pdfsubject={\papersubject},hidelinks]{hyperref}

\begin{document}

\maketitle
%
\begin{abstract}
Query-by-Humming (QbH) is a task that involves finding the most relevant song based on a hummed or sung fragment. Despite recent successful commercial solutions, implementing QbH systems remains challenging due to the lack of high-quality datasets for training machine learning models. In this paper, we propose a deep learning data collection technique and introduce Covers and Hummings Aligned Dataset (CHAD), a novel dataset that contains 18 hours of short music fragments, paired with time-aligned hummed versions. To expand our dataset, we employ a semi-supervised model training pipeline that leverages the QbH task as a specialized case of cover song identification (CSI) task. Starting with a model trained on the initial dataset, we iteratively collect groups of fragments of cover versions of the same song and retrain the model on the extended data. Using this pipeline, we collect over 308 hours of additional music fragments, paired with time-aligned cover versions. The final model is successfully applied to the QbH task and achieves competitive results on benchmark datasets. Our study shows that the proposed dataset and training pipeline can effectively facilitate the implementation of QbH systems.
\end{abstract}
\section{Introduction}\label{sec:introduction}
Query-by-Humming (QbH) is a well-known task in Music Information Retrieval. It aims to enable users to find a particular song within a retrieval system by providing a small audio segment of their voice or humming as a query. Such systems rely on a large database of songs and display the most similar matches to the user's query.

One significant benefit of the QbH system compared to other music search systems \cite{shazam} is that users do not have to play a copy of the song or recall its lyrics. Instead, they can hum or sing the melody of the desired song, and the system will use advanced audio processing and deep learning techniques to locate it.

A similar task to QbH is Cover Song Identification (CSI) task \cite{csi_1, csi_2, csi_3}. CSI aims to identify cover songs performed by different artists as versions of original songs within a music database. Although CSI systems often rely on neural networks, traditional QbH systems mainly utilize audio processing and music information retrieval techniques like pitch estimation, note extraction, and time series matching \cite{qbh4432653, qbh7352471, qbh1237790}. The main reason QbH lacks deep learning models is the absence of large datasets for training. This is primarily due to the high cost and limited availability of humming/singing data for QbH compared to CSI, where multiple versions of the same song are sufficient for training. Additionally, QbH requires the alignment of humming/singing fragments with the original versions of the song.

To overcome the challenges of limited data, we propose a 
novel dataset CHAD - Covers and Hummings Aligned Dataset. This dataset contains groups of time-aligned music fragments, primarily consisting of vocal segments from popular songs. \textit{Time alignment} is the process of synchronizing a fragment from a humming or cover version of a song with its corresponding fragment from the original version to have the same temporal structure. The groups are separated into two categories: one with humming fragments collected via crowdsourcing and another with cover fragments collected using a semi-supervised training pipeline. We use this dataset to train our deep learning model for matching audio fragments with similar melodies using metric learning paradigm. We demonstrate that these techniques can also be successfully applied in the QbH task, achieving results comparable to the best performing scores on popular QbH benchmarks. Furthermore, we evaluate our model's performance on a large internal song database, showing its ability to generalize to a wider range of songs.

The paper is structured as follows. Section \ref{sec:Related} briefly reviews existing approaches to the QbH task. Section \ref{sec:Model} describes the proposed deep learning model and training method for the QbH task. Section \ref{sec:Data} outlines the dataset and semi-supervised data collection pipeline. Section \ref{sec:Experiments} describes the experiments conducted on public and private data. Finally, Section \ref{sec:Conclusions} concludes the paper.
\section{Related Work}
\label{sec:Related}
QbH systems typically have two components: audio transcription and search modules. Many approaches in QbH research have focused on designing effective representations of hummings that can be easily matched with MIDI targets. Some standard methods include using Hidden Markov Models \cite{qbh4432653} to transcribe hummings into a sequence of symbols, discretizing fundamental frequency into semitones \cite{qbh7352471}, and transcribing hummings into a note-like structure using pitch, interval, and duration features \cite{qbh1237790}.

Once the humming has been transcribed into a format that can be compared to database entries, the search module is responsible for finding the most relevant songs. Dynamic Time Warping \cite{qbh7352471, bs_qbh} has been a popular algorithm for comparing the humming query to MIDI-audio entries in the database. This algorithm finds the minimum path between the discretized humming and the MIDI-audio database. Another approach, top-down melody matching \cite{qbhWu2006ATA}, involves dynamically aligning the humming query with a song from the database. A third approach, progressive filtering \cite{qbh4432642}, involves multiple stages of song recognition with increasingly complex recognition mechanisms. These algorithms serve to match humming queries with songs in a database effectively. In the approach \cite{review_add1}, authors use melody extraction network to extract robust features from audios and match them with songs from database using an ensemble of melody matching algorithms.


In contrast to QbH, the latest research on the Cover Song Identification (CSI) task has been focused on deep learning-based techniques. A popular approach in CSI is to use deep neural networks for audio representation and metric learning for similarity search. In \cite{tiktok1}, the authors use Constant-Q Transform (CQT) of audio and train a modified version of ResNet with two losses - triplet loss for intra-class compactness and classification loss for inter-class discrimination. In the next study, \cite{tiktok2}, the authors improve results by integrating the PCA module into the fully-connected layer of ResNet. Metric learning is widely used in CSI. It is shown that different model and loss architectures like Siamese Network \cite{csi_2}, triplet loss \cite{tiktok1}, and contrastive loss \cite{csi_3} can produce competitive results.

In \cite{csi_1}, the authors use VGG on CQT features with variable length to tackle the problem of tempo changes of the cover songs. In \cite{tralie_1}, the authors use an audio signal's Mel-Frequency Cepstrum Coefficients (MFCC) as the representation. They build cross-similarity matrices between songs and collect the nearest neighbors of each song based on these matrices.

Several datasets are available for CSI tasks \cite{dataset_cover80, dataset_shs, dataset_kara1k, dataset_datacos}. These datasets contain audio features alongside music metadata and provide researchers with a way to evaluate and validate their models without collecting large amounts of audio data.
\section{Model}\label{sec:Model}
\begin{figure}[!t]
 \small
 \centering{
    \includegraphics[width=0.5\columnwidth]{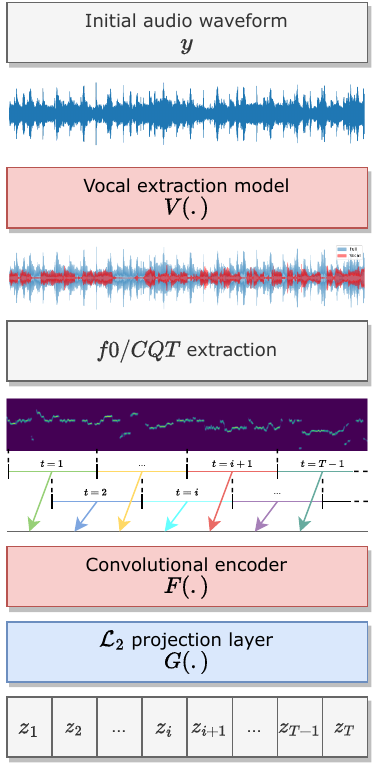}
 }
 \caption{Audio encoder model. Vocal part is extracted from the input waveform. Then, either $f0$ or $CQT$ features are calculated on the vocal part. Finally, the features are processed by a convolutional encoder model and, then, the output embeddings are normalized.}
 \label{fig:neural_fingerprint}
\end{figure}
Our encoder model, $M$, is presented in Figure \ref{fig:neural_fingerprint} and inspired by \cite{csi_3}. The whole fingerprints extraction pipeline can be described in the following steps:
\begin{enumerate}
    \item The first step of the encoding process is to extract the vocal part of the audio waveform $y$ using a pre-trained audio source separation model $V(.)$. The model is applied only to cover fragments since humming fragments do not contain any accompaniments. We used Spleeter \cite{spleeter2020} as a model due to its high-speed performance.
    \item The vocal part of the audio is then sent to the feature extractor model. In this study, two different extracted feature types are used: the first is the fundamental frequency ($f0$) extracted using the CREPE model \cite{crepe}, which is considered a robust representation of the melody. The second is Constant-Q Transform (CQT) as its faster alternative. The melody is crucial for search as it contains essential song information while ignoring irrelevant singing person details.
    \item The extracted feature matrix is then separated into overlapping segments, called analysis windows with length $W$ and step $H$, which are fed separately to a convolutional encoder $F(.)$ ResNet18 \cite{resnet}. The final layer of the encoder is a $\mathcal{L}_2$-normalization layer $G(.)$, which normalizes the output of the encoder along the embedding dimension.
    \item The output fingerprints $Z=\{z_i\}_{i=1...T}$, where $T$ is the total number of fingerprints for a waveform and 128 is its dimension size.
\end{enumerate} 

We use the metric learning method similar to \cite{csi_3} as a learning framework. To form a batch of audio fragments for training, we randomly sample $K$  groups of time-aligned audio fragments. By \textit{group}, we refer to a collection of original song and humming/singing fragments. Then, we select $n$ random audio fragments from each group and extract a random analysis window of size $W$. Since our data is aligned, all windows from each group will represent the variations of the same data. Afterward, we apply our model and extract in total $N=K\cdot n$ fingerprints for $n$ in each group. 

Our loss is defined as follows:
\begin{equation}
    \ell = -\sum_{k=1}^K \sum_{z_k^i, z_k^j \in Z_k} \log \frac{\exp(\frac{sim(z_k^i,z_k^j)}{\tau})}{\sum_{z_l \not\in Z_{k}}\exp(\frac{sim(z_k^i,z_l)}{\tau})},
\end{equation}
where $Z_k=\{z_k^0,\dots z_k^{n-1}\}$ is the group of fingerprints, $z_k^i$ and $z_k^j$ are different fingerprints from $Z_k$, $z_l \not\in Z_{k}$ stands for all fingerprints not in a given group $k$,  $sim(z_i, z_j)=z_i^T z_j$ is the similarity function, and $\tau$ is a temperature parameter.  The final loss is computed across all possible positive pairs and averaged afterward.
\section{Dataset}\label{sec:Data} 

\begin{figure}[t!]
 \centering{
 \includegraphics[width=0.65\columnwidth]{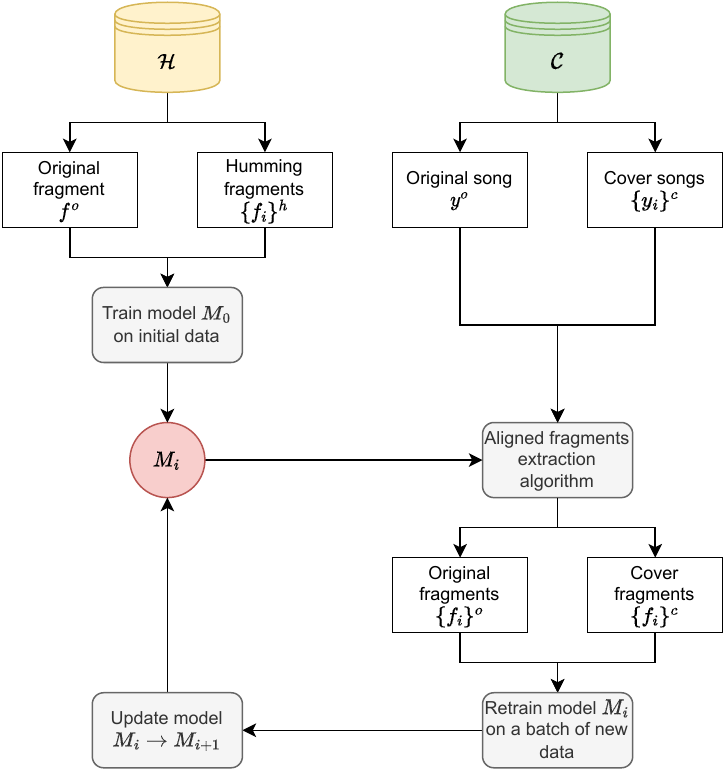}
 }
 \caption{Semi-supervised training and data collection pipeline used to train the initial model, and iteratively gather new aligned audio fragments and retrain the model.}
 \label{fig:semisup_pipeline}
\end{figure}

This section describes the process of collecting a dataset for the QbH task, its statistics, and its limitations.

\subsection{Semi-supervised pipeline}

\begin{algorithm}[t!]\footnotesize
    \SetKwInOut{Input}{Input}
    \SetKwInOut{Output}{Output}
    \Input{
    $y^o$ - original song;\\ 
    $Y^c$ - set of cover songs;\\ 
    $d_{min}$ - fragment's minimal length;\\
    $d_{max}$ - fragment's maximal length;\\
    $D_{p}$ - set of pause lengths;\\ 
    $L_{db}$ - set of dB levels;\\ 
    $\alpha_{corr}$ - threshold value to exclude same fragments;\\
    }
    \Output{
    $\mathbb{F}$ - set of groups of aligned fragments from original and cover songs
    }
    $\mathbb{F}^o, \mathbb{F}^o_{prev} \leftarrow \{\}, \{\}$ \;            \label{alg:step1}
    $M \leftarrow$ initialize\_model$(.)$\;
    $rms \leftarrow $ rms$(y^o)$\;
    \ForEach{$d_{p} \in D_{p}$}{                                        \label{alg:step2}
        \ForEach{$l_{db} \in L_{db}$}{
            $m_{silence} \leftarrow$ find\_silence\_mask$(rms, l_{db})$\;   \label{alg:step3}
            $\mathbb{F}^o \leftarrow$ split\_by\_silence$(y^o, m_{silence})$\;
            $\mathbb{F}^o \leftarrow$ merge\_fragments$(\mathbb{F}^o, d_{p}, d_{min}, d_{max})$\;
            $\mathbb{F}^o_{emb} \leftarrow M(\mathbb{F}^o)$\;             \label{alg:step4}
            $\mathcal{A}_{corr} \leftarrow$ build\_correlation\_matrix$(\mathbb{F}^o_{emb})$\;
            $\mathbb{F}^o \leftarrow$ find\_unique\_fragments$(\mathbb{F}^o, \mathcal{A}_{corr}, \alpha_{corr})$\;
            $\mathbb{F}^o, \mathbb{F}^o_{prev} \leftarrow max(\mathbb{F}^o, \mathbb{F}^o_{prev})$\; \label{alg:step5}
        }
    }
    $\mathbb{F} \leftarrow \{\}$\; \label{alg:step6}
    \ForEach{$f^{o} \in \mathbb{F}^o$}{ 
        \ForEach{$y^{c} \in \mathbb{Y}^c$}{ 
            $f^{o}_{emb} \leftarrow M(f^o)$\; \label{alg:step7}
            $y^{c}_{emb} \leftarrow M(y^c)$\;
            $\mathbb{F}^c \leftarrow$ cross\_correlation($y^c, y^{c}_{emb}, f^{o}_{emb}$)\; \label{alg:step8}
            $\mathbb{F}^c \leftarrow$ filter($\mathbb{F}^c, \beta_{rel}, \beta_{irrel}$)\;
            $\mathbb{F} \leftarrow \mathbb{F} \cup (f^{o}, \mathbb{F}^c)$
        }
    }
    
\caption{Aligned fragments extraction algorithm of data collection pipeline.}
\label{alg:alg1}
\end{algorithm} 

Figure \ref{fig:semisup_pipeline} presents our proposed semi-supervised pipeline. The dataset used in this study is structurally divided into two parts: $\mathcal{H}$ and $\mathcal{C}$. The first part, $\mathcal{H}$, consists of original music fragments $f^o$ paired with time-aligned humming/singing fragments $f^h$, making groups $F^h$. The $f^o$ fragments are represented by various vocal and instrumental parts of music clips. The $f^h$ fragments were collected using a crowdsourcing service Yandex.Toloka. The second part of the dataset, $\mathcal{C}$, was created by collecting the 100 most popular songs from the Billboard Charts for each year from 1960 to 2020. 
For each song, up to 10 cover versions were retrieved from YouTube top results using query "\{song name\} \{artist name\} cover".

Because $\mathcal{H}$ already has groups of time-aligned fragments, we can train the initial encoder model $M_0$ with this data. However, $\mathcal{C}$ only contains groups of full song versions instead of time-aligned fragments, so extracting fragments from these groups is necessary. We propose Algorithm \ref{alg:alg1} for this task. This algorithm is designed to extract the maximum amount of unique fragments from the original versions of the songs and find the corresponding aligned fragments from cover versions of the songs in $\mathcal{C}$. The algorithm can be described in three stages:  \\\\
\noindent \textbf{Initialization stage}
\begin{enumerate}
    \item As input, the algorithm takes the vocal part of the original song $y^o$ and a group of cover songs $Y^c$. Additionally, the algorithm takes the minimal and maximal length of the fragment $d_{min}$ and $d_{max}$, respectively, the set of dB levels $L_{db}$ by which to count the region in song as silent or non-silent, the set of maximal pause lengths $D_{p}$ between adjacent fragments in a song separated by silence to be considered as one fragment, and threshold values $\alpha_{corr}$, $\beta_{rel}$, and $\beta_{irrel}$ to exclude unwanted fragments from the output set.
    \item Initialize empty sets of unique fragments $\mathbb{F}^o$ and $\mathbb{F}^o_{prev}$, the encoder model $M$, and $rms$ of the waveform $y^o$ (lines \ref{alg:step1}-3).
\end{enumerate}
\noindent \textbf{Fragmentation stage}
\begin{enumerate}
    \item To find the best combination of $D_{p}$ and $L_{db}$ to yield $\mathbb{F}^o$ of maximal size, start two loops by iterating over these sets (lines \ref{alg:step2}-5).
    \item Compute the binary mask of non-silent regions $m_{silence}$ using $rms$ and $l_{db} \in L_{db}$ and find a set of initial fragments by splitting the waveform $y^o$ using this mask. Then, merge adjacent fragments, the pause between which is less than $d_{p} \in D_{p}$. Additionally, the length of such fragments should satisfy the condition $d_{min}<|f|<d_{max}, f \in \mathbb{F}^o$. (lines \ref{alg:step3}-8).
    \item Apply model $M_i$ to the found fragments and extract the fingerprints. Then, build the correlation matrix $\mathcal{A}_{corr}$ based on the fragments' fingerprints $\mathbb{F}^o_{emb}$ and exclude the ones with a correlation higher than the threshold $\alpha_{corr}$. We used the maximum of cross-correlation function to measure the correlation of fingerprints with different lengths (lines \ref{alg:step4}-11).
    \item  Find the parameters of dB levels and pause lengths that yield the maximum amount of unique fragments (line \ref{alg:step5}).
\end{enumerate}
\noindent \textbf{Matching stage}
\begin{enumerate}
    \item Once the unique fragments from the original version of the song $\mathbb{F}^o$ are extracted, initialize the empty set $\mathbb{F}$ to be filled with groups of the time-aligned fragments from original and cover songs and iterate over each found original fragment $f^o \in {\mathbb{F}^o}$ and each cover song $y^c \in Y_{c}$ (lines \ref{alg:step6}-17). 
    \item Extract fingerprints from original fragment $f^o_{emb}$ and cover song $y^c_{emb}$ using $M$. Search for the same fragments in the cover song using a cross-correlation function and peak detection algorithm (lines \ref{alg:step7}-19). 
    \item Filter out noise cover fragments by establishing two thresholds:
    \begin{enumerate}
        \item The cover fragments with correlation above $\beta_{rel}$ are considered relevant, indicating a high level of certainty that the content of the cover fragment is similar to that of the original fragment. 
        \item The cover fragments with correlation below $\beta_{irrel}$ are considered irrelevant fragments and are excluded. Fragments with a correlation between these two thresholds are counted as uncertain and require double-checking via additional crowdsourcing.  
    \end{enumerate}
    Save the gathered groups of aligned fragments (lines \ref{alg:step8}-22).
\end{enumerate}

We apply this pipeline to song batches of $\mathcal{C}$, which generates new groups of aligned data. These groups are then added to $\mathcal{H}$, and the model, $M$, is retrained on the newly gathered data. In such a way, we first train $M_0$ on initial humming data, then iteratively update our model from $M_i \rightarrow M_{i+1}$ and fill our dataset with new data.

For the unique fragments extraction algorithm, we set $d_{min}=8$, $d_{max}=20$, $D_{p}=\{0.5, 1, 1.5\}$ seconds, $L_{db}=\{52, 56, 60, 64, 68\}$ dB, $\alpha_{corr}=0.8$. When searching for fragments in cover versions of songs, we set the optimal thresholds to $\beta_{rel}=0.5$ and $\beta_{irrel}=0.3$. All fragments from the same group have equal duration to retain the time-alignment consistency. 

\subsection{Statistics}

We call the collected dataset Covers and Hummings Aligned Dataset (CHAD). Here are the dataset's statistics:
\begin{itemize}
    \item CHAD contains $5494$ original songs, $31630$ cover songs, and $5164$ hummings fragments.
    \item The total number of audio fragments is $81781$, which amounts to over $270$ hours of singing/humming audio fragments and $51$ hours of original song fragments. The group size varies from 2 to 31, with an average size of 6 fragments.
    \item In $\mathcal{H}$, the duration of the fragments  ranges from $4$ to $20$ seconds, with a mean of $11.06\pm2.67$ seconds, and a total for original fragments - 2.12 hours, and for humming fragments - 15.83 hours.
    \item In $\mathcal{C}$, the duration ranges from $8$ to $20$ seconds, with a mean of $14.66\pm2.03$ seconds, and a total for original fragments - 49.54 hours, and cover fragments - 259.03 hours, where 194.53 hours are for fragments with correlation above $\beta_{rel}$, and 64.50 hours are for fragments with correlation between $\beta_{rel}$ and $\beta_{irrel}$.
    \item Additionally, the metadata is collected. It includes YouTube video ID, title, author, correlation value, and whether the fragment is double-checked. The dataset's audio IDs, metadata, start and end timestamps and data download script are available in our GitHub repository\footnote{\href{https://github.com/amanteur/CHAD}{https://github.com/amanteur/CHAD}}.
\end{itemize} 

\subsection{Limitations}
However, our semi-supervised pipeline has some limitations. First, it can only extract vocal data, and the algorithm needs modification to extract instrumental segments. Second, the number of covers is limited, as there are usually fewer cover versions for non-popular songs. Lastly, there will still be some noisy unrelated fragments in the final set due to the automatic validation threshold. Future research could explore using generative networks \cite{hum2search} to overcome these limitations.

\section{Experiments}
\label{sec:Experiments}
\begin{table*}[t!]
    \centering
    \begin{tabular}{cccccccc}
        \toprule
        \multirow{2}{*}{} & \multirow{2}{*}{Method}  & \multicolumn{5}{c}{Top-$10$ hit rate $\uparrow$} \\
        \cmidrule(lr){3-7}
         & & Jang\cite{dataset_jang} & Thinkit & Subtask 2 & Jang Real & MTG-QBH \cite{dataset_mtg} \\
        \midrule
        \multirow{2}{*}{Ours} & metric learning(CREPE) & 0.921 & 0.966 & 0.959 & 0.868 & 0.883 \\
        & metric learning($CQT$) & 0.840 & 0.786 & 0.866 & 0.867 & 0.747 \\
    	\midrule
        Stasiak \cite{bs_qbh} & $f0$-matching & 0.948 & 0.907 & 0.968 & - & - \\
        ACRCloud & proprietary  & \textbf{0.990} & \textbf{0.986} & \textbf{0.972} & - & - \\
    	\bottomrule
    \end{tabular}
    \caption{Evaluation of model $M_{short}$ trained on dataset $\mathcal{C} + \mathcal{H}$ with two types of features - CREPE and $CQT$. Evaluation is provided on MIREX datasets - Jang, Thinkit, and Subtask 2, and datasets - Jang Real, and MTG-QBH, which are more applicable to real-world scenarios. 
    }
    \label{tbl:results_MIREX_midi}
\end{table*}
\subsection{Experimental setup}
\textbf{Input features}. We used CREPE \cite{crepe} activations as $f0$ features, yielding output features with a size of $(360, T)$. We further enhanced the robustness of the melody feature by trimming it to include only $3$ octaves around its mean pitch, following the approach used in \cite{f0est_csi}. Additionally, we downscaled this representation to the size $(80, \frac{T}{4})$. However, we encountered issues with the slow speed of the melody extraction model during evaluation, rendering the overall approach unscalable. To address this, we incorporated $CQT$ features into our model, extracted with the following parameters: $12$ bins per octave with a total of $7$ octaves, Hann window, hop length $512$, and a sampling rate of $16$ kHz.

 \noindent \textbf{Augmentations}. We found an optimal set of augmentations to every batch of waveform fragments, which included continuous pitch shifting (with a shift range of -4.0 to 4.0 semitones and probability of 0.5), time stretching (with a stretch rate range of 0.8 to 1.25 and probability of 0.8), SpliceOut \cite{aug_spliceout} (with 10 random intervals of 500 frames and probability of 0.8), mixing with other audio samples in the batch (with an SNR range of 5 to 10 dB and probability of 0.8), and adding background noises (with an SNR range of 3 to 30 dB and probability of 0.8).

\noindent \textbf{Model}. We discovered that as the length of a hummed or sung recording increases, the tempo/rhythm becomes more mismatched from the original song. So we trained two models with different analysis window lengths ($W$) and hop sizes ($S$): $M_{short}$ for shorter recordings (up to $15$ sec) with $W$=$3$ sec and $S$=$0.25$ sec, and $M_{long}$ for longer recordings with $W$=$8$ sec and $S$=$0.64$ sec. Both models used a vanilla ResNet18 encoder model, with output embeddings of size $(128, T)$, where $T$ is the number of fingerprints.

\noindent \textbf{Training setup}. We trained the encoder model using the ADAM optimizer, with a learning rate of $lr=0.001$ and a batch size of $32$ for $100$ epochs. We used the NT-Xent Loss \cite{ntxent_loss} with a temperature of $t=0.05$. We employed the Multi-similarity miner \cite{multisimilarity} and an adaptive batch sampler to improve convergence speed. The batch sampler selects up to $4$ fragments with a random starting point for each fragments group. We trained the model under two settings: only on the $\mathcal{C}$ part and on both $\mathcal{C} + \mathcal{H}$ parts of CHAD. Models were trained on 1 NVIDIA GeForce RTX 2080 Ti 12 Gb. 

To evaluate the performance of our model, we conducted a series of experiments, which involved:
\begin{itemize}
    \item Experiments on the MIREX QbH datasets \cite{mirex_competition}, specifically the Roger Jang and ThinkIt datasets, where MIDI recordings were used as references. The MIR-QBSH corpus of Roger Jang consists of 4431 query hummings and 48 original MIDI files, while the Thinkit corpus contains 355 queries and 106 original MIDI files. The song database was constructed according to MIREX QbH Challenge standards, with 2600 MIDI files. These experiments aimed to find the ground-truth MIDI by a given query humming.
    \item Experiments on MIREX QbH datasets according to Subtask 2 testing protocol in MIREX evaluation system. In this protocol, queries are also considered as "versions" of ground truth, and the objective is to retrieve all variants related to a searched ground truth by given query humming.
    \item Experiments on a dataset of real recordings, which included MIREX Roger Jang Dataset with all MIDI files replaced with real recordings extracted from YouTube videos (Jang Real), and MTG-QbH \cite{dataset_mtg} dataset with $118$ queries and $118$ original songs. The additional database comprises 1886 random songs from the internal dataset to serve as imposter songs. 
    \item Experiments on a large-scale internal database (DB90K) containing more than 90k real song recordings. For this experiment, we used two types of queries: $126$ humming fragments for $126$ songs collected by our team as search-by-humming setup and $2000$ singing fragments from karaoke recordings gathered from the DAMP-VPB dataset \cite{DAMP} as search-by-singing setup. In the latter case, we selected $5$ of $16$ original songs and their sung performances and manually split them into fragments.
\end{itemize}

For all real recordings, we extract the vocal part beforehand. Also, we ensure that CHAD does not contain any songs that are also present in the evaluation datasets. This was achieved by excluding such songs from the training set.

\noindent \textbf{Retrieval}. We use two variants of sequence matching methods at the retrieval phase: maximum Pearson correlation coefficient (Corr) or Dynamic Time Warping (DTW) \cite{dtw_muller}. For the large-scale experiment on DB90K, we use a two-step search procedure with a first step of fast retrieval of preliminary candidates using the FAISS Approximate Nearest Neighbors (ANN) algorithm with Euclidean distance followed by a second step of reranking. After further analysis, we discovered that Euclidean distance and Cosine distance yielded similar results. To maintain simplicity, we chose to use Euclidean distance. The ANN search returns the top $5000$ candidates, which are reranked based on the Pearson correlation score.

\noindent \textbf{Metrics}. We follow the MIREX evaluation protocols \cite{mirex_competition} for the QbH task and compute the mean of the Top-$n$ hit rate for every humming/singing fragment. There was only one related song in the database for every query fragment. 

\subsection{Results}
We compare our model $M_{short}$ trained on $\mathcal{C} + \mathcal{H}$ with 2 best performing methods according to the latest available result of MIREX QbH Challenge \cite{mirex_results}. The first one \cite{bs_qbh} is based on $f0$-matching technique. The second one is a proprietary method for which only scores were reported in the leaderboard.  We use only one of our models ($M_{short}$ on CREPE and CQT features) in this experiment as most query fragments are shorter than 16 seconds. We use DTW for matching feature sequences on MIDI-based datasets and Corr for non-MIDI datasets as we found that correlation coefficient gives performance improvement on real data.

The results are summarized in Table \ref{tbl:results_MIREX_midi}. Our model demonstrates competitive though slightly inferior performance on the given benchmarks. On real (non-MIDI) data our implementation of \cite{bs_qbh} produced near-random results which can be explained by the difficulty of tracking and matching $f0$ in real music recordings. Also, we see that while using $CQT$ features led to a performance drop, it is not prohibitively large so $CQT$ features can be used when computing $f0$ is infeasible.


Table \ref{tbl:results_90K} shows the scalability of our approach in the experiment with DB90k. We do not track the top-1 hit rate as the database contains several versions of the same song. Table \ref{tbl:results_90K_humming} reports the results of the search-by-humming setup. We used $M_{short}$, $M_{long}$, and their combination model $M_{fused}$, which worked on a simple rule: $M_{short}$ was used for hummings shorter than 15 seconds, while $M_{long}$ was used otherwise. All presented models are trained with $CQT$ features. We observed that $M_{fused}$ worked better than $M_{short}$ and $M_{long}$ separately in all scenarios. Comparing models trained on $\mathcal{C}$ and $\mathcal{C}+\mathcal{H}$, the accuracy gap suggests that training on real humming data is crucial for search-by-humming setup. In Table \ref{tbl:results_90K_singing}, we report our results for search-by-singing setup with $M_{fused}$ on DAMP-VPB. Our model, trained on both $\mathcal{C}$ and $\mathcal{C} + \mathcal{H}$, retrieves the correct songs with high precision, with no performance drop observed for the model trained on $\mathcal{C}$ alone, due to the dominance of sung fragments in our training dataset. 

Additionally, we evaluate the retrieval speed of our models $M_{short}$ and $M_{long}$ on DB90k, as shown in Table \ref{tbl:results_90K_speed_test}. We find that $M_{long}$ performs better than $M_{short}$ in both search steps (ANN and Reranking) due to its ability to process longer humming recordings and thus require less processing of fragments. Our results demonstrate the scalability and efficiency of our search system in efficiently achieving high-precision results.

\begin{table}[t!]
    \small
    \centering
    \begin{subtable}[t]{\columnwidth}
    \centering
	\begin{tabular}{cccccc}
        \toprule
		\multirow{2}{*}{Partition} & \multirow{2}{*}{Model} & \multicolumn{4}{c}{Top-$n$ hit rate$\uparrow$}\\
		\cmidrule(lr){3-6}
		                           &                        & $100$    & $10$  & $5$  & $3$ \\
		\midrule
		\multirow{3}{*}{$\mathcal{C}$} & $M_{short}$                          & 0.643   & 0.548 & 0.524 & 0.476 \\
		                        & $M_{long}$                         & 0.412   & 0.277 & 0.270 & 0.262 \\
		                        & $M_{fused}$             & 0.759   & 0.621 & 0.603 & 0.517 \\
		\midrule
		\multirow{3}{*}{$\mathcal{C} + \mathcal{H}$} & $M_{short}$                          & 0.659   & 0.595 & 0.571 & 0.484 \\
		                        & $M_{long}$                          & 0.595   & 0.508 & 0.413 & 0.389 \\
		                        & $M_{fused}$             & 0.776   & 0.707 & 0.691 & 0.586 \\
        \bottomrule
	\end{tabular}
    \caption{Results on humming queries.}
	\label{tbl:results_90K_humming}
    \end{subtable}
    \par\medskip
    \begin{subtable}[t]{\columnwidth}
    \centering
	\begin{tabular}{*{6}{c}}
        \toprule
		\multirow{2}{*}{Partition} & \multirow{2}{*}{Model} & \multicolumn{4}{c}{Top-$n$ hit rate$\uparrow$}\\
		\cmidrule(lr){3-6}
		&                & $100$    & $10$  & $5$  & $3$ \\
		\midrule
		$\mathcal{C}$   & \multirow{2}{*}{$M_{fused}$}     & 0.931  & 0.904 & 0.885 & 0.865 \\
		$\mathcal{C} + \mathcal{H}$ &      & 0.923   & 0.899 & 0.885 & 0.856 \\
        \bottomrule
	\end{tabular}
    \caption{Results on singing queries.}
    \label{tbl:results_90K_singing}
    \end{subtable}
    \caption{Evaluation on DB90K with humming and singing fragments using models  $M_{short}$, $M_{long}$, and their fusion model $M_{fused}$ trained on $\mathcal{C}$ and $\mathcal{C} + \mathcal{H}$ with $CQT$ features.}
    \label{tbl:results_90K}
\end{table}

\begin{table}[t!]
    \centering
    \begin{tabular}{*{3}{c}}
        \toprule
		\multirow{2}{*}{Model} & \multicolumn{2}{c}{Search step, s}\\
		\cmidrule(lr){2-3}
		                       & ANN               & Reranking \\
		\midrule
		$M_{short}$                 & 1.41 $\pm$ 0.57   & 5.37 $\pm$ 0.87 \\
		$M_{long}$                 & 0.52 $\pm$ 0.11   & 2.39 $\pm$ 0.43 \\
        \bottomrule
	\end{tabular}
    \caption{Query search speed on DB90K using models  $M_{short}$ and $M_{long}$ trained on $\mathcal{C} + \mathcal{H}$ with $CQT$ features using 32 CPU.}
    \label{tbl:results_90K_speed_test}
\end{table}

\section{Conclusions}
\label{sec:Conclusions}
In this paper, we propose a novel dataset CHAD alongside a semi-supervised data collection and training pipeline for a Query-by-Humming system. We show that cover songs could be used to train query-by-humming models with competitive performance. Although the model trained on open data performs well on sung queries, the pure search-by-humming setup requires adding a portion of real humming data into the training set for acceptable performance. The main disadvantage of the proposed approach is that it cannot be used for searching instrumental tracks. One possible solution to this problem would lie in the field of dominant melody extraction and generative networks and is left for future research.

\bibliography{a_semi-supervised_dl_approach}

\end{document}